\documentclass[a4paper]{jpconf}

\usepackage{iopams}
\usepackage{graphicx}
\usepackage{enumerate}
\usepackage{tikz}

\newcommand{\beq}{\begin{equation}}
\newcommand{\eeq}{\end{equation}}
\newcommand{\beqn}{\begin{eqnarray}}
\newcommand{\eeqn}{\end{eqnarray}}
\newcommand{\Z}[1]{\mathbb{Z}_{#1}}
\newcommand{\nn}{\nonumber}
\newcommand{\bt}[1]{| #1 \rangle}

\def\JHEP#1#2#3{{\it JHEP} {\textbf{#1}} (#2) #3}

\def\NPB#1#2#3{{\it Nucl.\ Phys.}\/ {B \textbf{#1}} (#2) #3}
\def\PLB#1#2#3{{\it Phys.\ Lett.}\/ {B \textbf{#1}} (#2) #3}

\def\PRT#1#2#3{{\it Phys.\ Rep.}\/ {\textbf{#1}} (#2) #3}

\begin{document}
\title{Discrete symmetries in the heterotic-string landscape }

\author{P Athanasopoulos}

\address{Department of Mathematical Sciences,\\
University of Liverpool,\\
Peach Street,\\
Liverpool L69 7ZL,\\
UK }

\ead{panos@liverpool.ac.uk}

\begin{abstract}
We describe a new type of discrete symmetry that relates heterotic-string models. It is based on the spectral flow operator which normally acts within a general $N=(2,2)$ model and we use this operator to construct a map between $N=(2,0)$ models. The landscape of $N=(2,0)$ models is of particular interest among all heterotic-string models for two important reasons:
Firstly, $N=1$ spacetime SUSY requires $(2,0)$ superconformal invariance and
secondly, models with the well motivated by the Standard Model $SO(10)$ unification structure are of this type. This idea was inspired by a new discrete symmetry in the space of fermionic $\mathbb{Z}_2\times \mathbb{Z}_2$ heterotic-string models that exchanges the spinors and vectors of the $SO(10)$ GUT group, dubbed spinor-vector duality. We will describe how to generalize this to arbitrary internal rational Conformal Field Theories.
\end{abstract}

\section{Introduction}
String theory provides the most promising framework for a fundamental theory of physics. Many semi-realistic string models have been constructed and understanding how many models that closely resemble the SM (or MSSM) are present is an extremely interesting question. Ultimately, this would potentially help find a dynamical way to select between these models (vacua).

In this context, the space of the heterotic $N=(2,0)$ models naturally draws the attention. These are models with spacetime supersymmetry (SUSY) which are also ``minimal", in the sense that $N=1$ spacetime SUSY requires (at least) $(2,0)$ superconformal invariance \cite{WeakAngle}. Furthermore, these models can accommodate an $SO(10)$ gauge group which is the most appealing candidate in grand unified theories (GUTs). 

The landscape of $(2,0)$ models is not nearly as well understood as in the $(2,2)$ case. As such, finding landscape symmetries can provide important insight towards that goal. The most famous example of such a symmetry is of course \emph{mirror symmetry} \cite{Mirror1}\footnote{See also \cite{Blumenhagen1997} for mirror symmetry in $(2,0)$ models which we are interested in here, and \cite{FracMirror} for a discussion similar in spirit to ours in the context of type II strings.}. The \emph{spinor-vector duality} and its extension discussed here is another discrete symmetry of the same type. 

The spinor-vector duality \cite{svd}  exchanges the spinors and vectors of the $SO(10)$ GUT group and applies to a particularly interesting sub-class of the above models, namely  the fermionic $\mathbb{Z}_2\times \mathbb{Z}_2$ ones. In this talk, which is based on results published in \cite{Me}, I will describe how to extend this idea beyond  $\mathbb{Z}_2\times \mathbb{Z}_2$ models, generalizing it to models with arbitrary internal rational Conformal Field Theories (CFTs).

\section{Notations and conventions}
We are discussing the heterotic string in  which the left-moving sector is supersymmetric and the right-moving sector is bosonic. A general state in a heterotic model is of the form:

\beq
\Phi_L\otimes\Phi_R
\eeq

with

\beq
\Phi_L=(w_L)(h_L,Q_L),\quad \Phi_R=(w)(h,Q)(p).
\eeq

Furthermore
\begin{itemize}
\item $w_L$ is an $SO(2)$ weight $(o, v, s, c)$
\item $w$ is an $SO(10)$ weight $(o, v, s, c)$
\item $p$ an $E_8$ weight.
\end{itemize}
We have used the common abbreviations for the four conjugacy classes (scalar ($o$), vector ($v$), spinor ($s$), conjugate spinor ($c$)) of $SO(2n)$. The numbers $h$ and $Q$ are the internal CFT components of the state and are respectively known as the \emph{conformal dimension} and the \emph{$U(1)$ charge}.

The appearance of the $SO(10)$ and $E_8$ weights is very general in heterotic models and can be thought of as arising from the \emph{bosonic string map} \cite{Gepner,BosonicStringMap}.
The bosonic string map provides an elegant way to make compatible the following statements:
\begin{enumerate}[i)]
\item The only generic means of achieving modular invariance in a non-free CFT is to have a left-right symmetric spectrum.
\item In the heterotic string, by definition, the left and right sectors are treated differently.
\end{enumerate}
\medskip

The recipe then is to construct heterotic models from type II models using the bosonic string map that effectively replaces the weights on the right-moving sector in the following way
$$w_{SO(2)}\rightarrow (w_{SO(10)}, w_{E_8})$$
This preserves the modular invariance of the theory, giving a well defined heterotic model.

\section{$(2,0)$ models}\label{sec:20models}
When referring to $(2,0)$ models, the numbers specify the amount of worldsheet supersymmetry in the left and right-moving sectors respectively.

By definition a CFT is said to have $N=2$ worldsheet supersymmetry if it includes four fields:
\beqn
T(z)&=&\sum_{n\in\Z{}}L_nz^{-n-2}\ ,\\
G^{\pm}(z)&=&\sum_{n\in\Z{}}G^{\pm}_{n\pm a}z^{-n-\frac{3}{2}\mp a}\ ,\\
J(z)&=&\sum_{n\in\Z{}}J_nz^{-n-1}\ ,
\eeqn
that satisfy the algebra \cite{ChiralRings}:
\beqn
\left[L_m,L_n\right]&=&(m-n)L_{m+n}+\frac{c}{12}(m^3-m)\delta_{m+n,0}\ ,\nn\\
\left[L_m, G^\pm_{n\pm a}\right]&=&\Big(\frac{m}{2}-n\mp a\Big)G^\pm_{m+n\pm a}\ ,\nn\\
\left[L_m, J_n\right]&=& -n J_{m+n}\ ,\nn\\
\left[J_m, J_n\right]&=& \frac{c}{3}m\delta_{m+n,0}\ ,\nn\\
\left[J_m, G^\pm_{n\pm a}\right]&=&\pm G^\pm_{m+n\pm a}\ ,\nn\\
\{G^+_{m+a},G^-_{n-a}\}&=&2 L_{m+n}+(m-n+2a)J_{m+n}+\frac{c}{3}\Big((m+a)^2-\frac{1}{4}\Big)\delta_{m+n,0}\ ,\nn\\
\{G^+_{m+a},G^+_{n+a}\}&=&\{G^-_{m-a},G^-_{n-a}\}=0\ ,\label{eq:algebra}
\eeqn

The stress-energy tensor $T$ is defined for any CFT. The requirement of $N=2$ worldsheet SUSY means that its superpartners should also be present. $G^{\pm}$ are the fermionic superpartners and $J$ is a $U(1)$ current.

The algebra also includes the real continuous parameter $a$ which describes how the fermionic superpartners $G^{\pm}$ of $T$ transform:
\beq
G^{\pm}(e^{2\pi i}z)=-e^{\mp2\pi i a}G^{\pm}(z).
\eeq
The algebras for $a$ and $a+1$ are actually isomorphic, with $a\in\Z{}$ corresponding to the R sector and $a\in\Z{}+\frac1 2$ corresponding to the NS sector. A state of the theory is completely described by the conformal dimension $h$ and the $U(1)$ charge $Q$ of the operators $L_0$ and $J_0$ that form the Cartan  subalgebra:
\beq
\bt{\phi}=\bt{h,Q}.
\eeq

One of the most interesting features of the algebra (\ref{eq:algebra}) which provides it with a rich structure is the fact that it is invariant under the following transformation:
\beqn
L_n^\eta&=&L_n+\eta J_n+\frac{c}{6}\eta^2\delta_{n,0}\ ,\nn\\
G^{\eta \pm}_{n\pm a}&=& G^{\eta \pm}_{n\pm(a+\eta)}\ ,\nn\\
J^\eta_n&=&J_n+\frac{c}{3}\eta\delta_{n,0}.
\eeqn
 which is known as the \emph{spectral flow}. Here $\eta$ is also a continuous real parameter.
This allows the definition of a \emph{spectral flow operator} $U_\eta$ that acts on states in the following way:
\beq
U_\eta\bt{h,Q}=\bt{h_\eta,Q_\eta}={\big| h-\eta Q+\frac{\eta^2 c}{6},Q-\frac{c \eta}{3} \big\rangle}
\eeq
and which will play a crucial role in what follows. Intuitively, the spectral flow operator is there because every supersymmetric theory must have an operator that relates the superpartners. On the left-moving sector, the spectral flow operator is just the supercharge and relates spacetime bosons to spacetime fermions. On the right-moving sector, the same operator relates spinorial representations of $SO(10)$ to vectorial representations of $SO(10)$. 

An amusing fact to note is that for $(2,2)$ models we have $N=2$ worldsheet SUSY even though there are no fermions on the worldsheet and all the degrees of freedom are described by bosons! This is feasible because of the equivalence of bosons and fermions on a 2d CFT.

One of the reasons that $(2,0)$ models are good candidates for string models which are phenomenologically viable is their (semi-)realistic gauge group. They arise from $(2,2)$ models  when the superconformal symmetry in the right-moving sector is broken. $(2,2)$ models have (at least) $E_6$ gauge symmetry which is phenomenologically unacceptable. This problem is largely rectified by considering $(2,0)$ models because the breaking of the superconformal symmetry is tied to the breaking of the gauge group.

The gauge group could in principle break to any of its subgroups, but one of the most well studied cases is having a remaining $SO(10)$ symmetry. The standard model particles fit nicely into representations of $SO(10)$ making such models prime examples of GUTs.

This breaking also lies at the heart of a novel symmetry known as \emph{spinor-vector duality}. When $E_6$ breaks to $SO(10)$ the matter representations decompose as:
\begin{eqnarray*}
\bf{27}&\rightarrow&\bf{16+10+1}\\
\bf{\overline{27}}&\rightarrow&\bf{\overline{16}+10+1}
\end{eqnarray*}
of which the $\bf{16}$ and $\bf{\overline{16}}$ are spinorials and the $\bf{10}$ vectorials.
It was noted in \cite{svd} that if the symmetry is broken in a particular way, namely using $\Z2$ or $\Z2\times\Z2$, then we would naturally end up with two different models: In one of them the spinorials would remain in the spectrum with the vectorials and singlets projected out and in the other one the situation would be reversed.

Even though the two models will not be physically equivalent, this is a very interesting symmetry in the string-landscape which connects two different vacua and demands/implies the existence of one from the existence of the other. The extension of this idea beyond  $\mathbb{Z}_2\times \mathbb{Z}_2$ breakings of $E_6$ and the generalization to models with arbitrary internal rational CFTs was done in \cite{Me} and will be described shortly, but first we need to introduce a few more tools.

\section{Simple currents}
The simple current method was developed by Schellekens and Yankielowicz in \cite{SimpleCurrents} (see also \cite{Kreuzer1994}) and provides a way to construct new modular invariants from a given one. It is similar in spirit to orbifolding but applies to a general rational CFT.

For a unitary rational CFT, the highest weight representations $[\phi_i]$ will satisfy a fusion algebra
\beq
[\phi_i]\times[\phi_j]=\sum_k N_{ij}^k [\phi_k],\quad N_{ij}^k\in\mathbb{N}_0.
\eeq
This expresses the fact that the product of two such states can be written as a linear combination using only states from within the same set.

\medskip

\noindent\textit{Definition:} A highest weight representation $\beta$ is called a \emph{simple current} if its fusion with any other highest weight
takes the form
$$[\beta]\times[\phi_i]=[\phi_{\beta(i)}].$$
Namely, for a simple current there is only a single state contributing to the RHS instead of a linear combination of many of them. Note that in the above definition the phrase ``simple current" should be treated as a single term. In particular, simple currents are not currents in the CFT sense.

\medskip

Futhermore, it is customary in many constructions to use additive notation for the states:
$$\beta+\phi_i=\phi_{\beta(i)}$$
and we adopt this from now on.

The simple current method is perhaps most readily understood by looking at the partition functions. Any string model will have a partition function
\beq
Z[\tau,\bar\tau]=\sum_{i,j} \chi_i(\tau)M_{ij}\chi_{j}(\bar\tau),
\eeq
where the holomorphic characters $ \chi_i(\tau)$ effectively encode the information about how many states there are on the left-moving sector with a particular $m_L^2$ and analogously for $\chi_{j}(\bar\tau)$ on the right-moving sector. The matrix $M$ in the middle  describes how to combine the two and is called a \emph{modular invariant}.

Starting with the model above and after selecting a simple current $\beta$, we can construct a new model with partition function
\beq
Z[\tau,\bar\tau]=\sum \chi_i(\tau)M_{ik}M_{kj}(\beta)\chi_{j}(\bar\tau),
\eeq
where
\beq\label{eqn:SCMI}
M_{kj}(\beta)=\frac 1 N\sum_{n=1}^{N_J}\delta({\Phi_k,\Phi_j+n\beta})\cdot \delta_{\mathbb{Z}}(Q_\beta(\Phi_k)+\frac{n}{2}Q_\beta(\beta))
\eeq
is called a \emph{simple current modular invariant}. The first delta function signifies the appearance of new (twisted with respect to $\beta$) sectors, whilst the second delta function implements the invariance projection. Namely only states with particular values of the \emph{monodromy charge} $Q_\beta$ will remain in the spectrum. The monodromy charge is defined as
\beq\label{eq:monodromycharge}
Q_\beta(\Phi)=h(\Phi)+h(\beta)-h(\beta+\Phi)\quad \mbox{mod }1.
\eeq
We can therefore see that in the new model some states have been projected out and some new states have appeared as well.

\begin{figure}
\tikzstyle{abstract}=[circle, thick, draw=black, rounded corners,
        text centered, anchor=center, text width=0.7cm]
\center\begin{tikzpicture}[scale=0.8,<->]
\node[abstract] (M) at (0,4) {$\mathcal{M}$\break (2,2)};
\node[abstract] (M0) at (6.5,6) {\centering{$\mathcal{M}_0$\break (2,0)}};
\node[abstract] (M1) at (6.5,4) {$\mathcal{M}_1$\break (2,0)};
\node (dots) at (6.5,2.5) {\textbf{\vdots}};
\node[abstract] (Mm) at (6.5,1) {$\mathcal{M}_m$\break (2,0)};
\path[] [->,double]
(M) edge node[above=1pt]{$\beta$} (M0);
\path[] [->,double]
(M) edge node[above=-3pt]{$\beta+ \beta_0$} (M1)
(M) edge node[below=5pt]{$\beta+m\beta_0$} (Mm);
\end{tikzpicture}
\caption{Diagram showing how one can start from a single $(2,2)$ model $\mathcal{M}$ and one simple current $\beta$ and construct many $(2,0)$ models. Reading it in reverse, many different  $(2,0)$ models can be associated with one $(2,2)$ parent model.}\label{fig:1}
\end{figure}
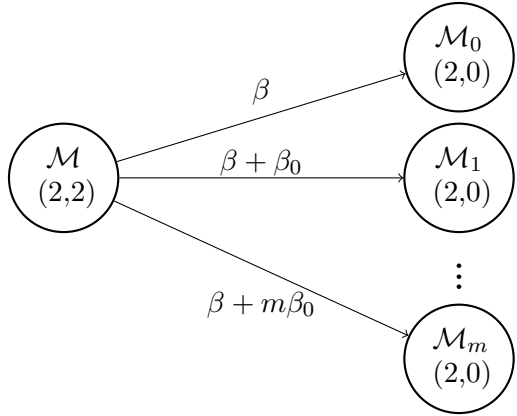

\section{The families of models}
We are now ready to describe the following simple, yet powerful idea, summarized in Figure \ref{fig:1}:

\begin{enumerate}[1.]
\item We use any $(2,2)$ model $\mathcal{M}$ as our starting point.
\item We choose a simple current $\beta$. Any $\beta$ would do, but the simple currents that we have in mind will break $E_6$ and give a $(2,0)$ model  $\mathcal{M}_0$.

\item As explained in section \ref{sec:20models}, since we started with a model with $N=2$ superconformal symmetry, we know that the spectral flow operator $\beta_0$ is another simple current that is present.

\item Using the fact that the sum of two simple currents is also a simple current, we can now consider the action of all simple currents $\beta+n\beta_0$ with $n=0,1,\ldots,N_{\beta_0}-1$ and generate many $(2,0)$ models from the same parent $\mathcal{M}$.
\end{enumerate}

The quantity $N_{\beta_0}$ appearing in step 4 above is called the \emph{order} of the spectral flow operator and is equal to the length of the orbit of $\beta_0$. For rational CFTs this is always a finite integer and the family of models obtained in this way is also finite.

One of the most natural questions that arises at this point is how the spectra of the models within the same family are related. Unsurprisingly enough, because of the complete generality of the construction we cannot possibly extract any specific spectra, but we can nevertheless derive useful relations among them. Even though the calculation that leads to them is fairly simple in principle, it does involve some technicalities and will be omitted here. The interested reader can find the details in \cite{Me}.

Let us outline some of the most interesting aspects of the results. Firstly, it turns out that the untwisted sectors in all the derived models are identical. In this context, untwisted sector means untwisted with respect to the simple current that defines the model, i.e states with $n=0$ in eqn (\ref{eqn:SCMI}). This points towards the direction that the models might be completely identical. The idea can be easily refuted by noting that the orders of $\beta+n\beta_0$ and  $\beta+m\beta_0$ will a priori be different, which means that these models will  in general have a different number of twisted sectors.

The interest then turns to examining the twisted sectors. It can be shown that massless states in the $n$-th twisted sector of the $m$-th model satisfy:

\beq
Q_{\beta}(\Phi_L)+\frac{n}{2}Q_{\beta}(\beta)+mn Q_{\beta_0}(\beta)\in\Z{},
\eeq

\beq\label{eq:cond2}
n\Big(h(\beta)+\frac{1}{2}Q_{\beta}(\beta)\Big)\in\Z{}.
\eeq

These relations encode a lot of information about where massless states can potentially be found and are especially useful when implementing computer scans over the spectra of such models. Note in particular, that eqn (\ref{eq:cond2}) is independent of the model $m$ in question! It states that for any model in the family, the only sectors $n$ we need to examine for massless states are those that make the quantity appearing in (\ref{eq:cond2}) an integer.

\section{Connection with spinor-vector duality and generalizations}
At this point, it is worth mentioning that simple currents of the form $\beta+n\beta_0$ with $n=0,1,\ldots,N_{\beta_0}-1$ are not the only ones that can be constructed from $\beta$ and $\beta_0$. We focused on this option because as it turns out the case with $N_{\beta_0}=2$ for which there are exactly two children models has already been studied. It is not hard to see that when restricting to the appropriate subspace of  $\mathbb{Z}_2$ and $\mathbb{Z}_2\times \mathbb{Z}_2$ models, the children models have the property that in one of them only massless vectorial states survive while in the other one only massless spinorial states survive. In other words, the above construction is a direct generalization of the spinor-vector duality.

However, we are free to venture even further and in fact it is natural to do so. A few ways to generalize the previous ideas would include the following:

\begin{enumerate}[i)]
\item Expanding the family generated by considering simple currents of the form  $n_1\beta+n_2\beta_0$ instead of just  $\beta+n\beta_0$.

\item Introducing discrete torsion $\epsilon$ between  $n_1\beta$ and $n_2\beta_0$.

\item Whenever the internal CFT can be written as a tensor product of $N=2$ superconformal theories, the Gepner models \cite{Gepner} being such an example, each term comes with a spectral flow operator $\beta_0^i$. We can then use only some of the $\beta_{0}^i$'s instead of the entire $\beta_{0}^{\mbox{\tiny{CFT}}}$.

\item Consider combinations of all the above.

\end{enumerate}

This seems to imply a great richness in the structure of the heterotic-string landscape, with the spinor-vector duality being only the tip of the iceberg. Even though the above possibilities have not been analyzed in detail, the methods of \cite{Me} do generalize straightforwardly to include these cases as well. This remains to be done in future work.

\section{Conclusions}
String theory is arguably the best candidate for a unified description of the four fundamental interactions of nature. In this talk, we focused on the heterotic-string landscape of $(2,0)$ models which are of great interest because of the requirement of spacetime SUSY and the accommodation of $SO(10)$ unification.

Discrete symmetries can be a very useful tool to understand the features of the landscape as they describe how different points/models are related. The most well known case of this type is mirror symmetry and a more recent example is the spinor-vector duality which was first observed in some classes of  $\mathbb{Z}_2$ and $\mathbb{Z}_2\times \mathbb{Z}_2$ orbifold models.

Upon generalizing this idea to theories with arbitrary rational internal CFTs using simple currents, we discovered that it is the spectral flow operator that in reality induces an entire family of models. This also  uncovered a vast structure of discrete symmetries in the landscape of $(2,0)$ vacua, the spinor-vector duality being only a particular case of the above.

We hope that studying these underlying symmetries of the heterotic-string landscape will prove useful in the long term goal of classifying completely all the $(2,0)$ models and maybe eventually help us come up with a dynamical vacuum selection mechanism in string theory.

\ack This talk is based on results derived in \cite{Me} in collaboration with Alon Faraggi and Doron Gepner, both of whom I am greatly indebted to. I would also like to thank the organizers of the conference for inviting me to give a talk. My work is supported by the Hellenic State Scholarships
Foundation (IKY).

\end{document}